\documentclass[doublecol]{epl2}
\usepackage{graphicx}
\usepackage{subfigure}
\usepackage{color}
\usepackage{hyperref}

\title{Superconductivity in multi-orbital $\mathrm{t-J_1-J_2}$ model and its implications for iron pnictides}

\author{Pallab Goswami \and Predrag Nikolic \and Qimiao Si}
\shortauthor{Pallab Goswami \etal}

\institute{
  \inst{1} Department of Physics and Astronomy, Rice University, Houston, TX 77005, USA
}

\abstract{
Motivated by the bad metal behavior of the iron pnictides,
we study a multi-orbital $\mathrm{t-J_1-J_2}$ model and
investigate possible singlet superconducting pairings.
Magnetic frustration by itself leads to a large degeneracy in the pairing
states. The kinetic energy breaks this into a quasi-degeneracy among
a reduced set of pairing states. For small electron and hole Fermi pockets,
an $A_{1g}$ state dominates over the phase diagram
but a $B_{1g}$ state has close-by energy.
In addition to the nodeless $A_{1g}$ $s_{x^2y^2}$ channel,
the nodal $A_{1g}$ $s_{x^2+y^2}$ and
$B_{1g}$ $d_{x^2-y^2}$ channels are also competitive in the
magnetically frustrated $J_1 \sim J_2$ parameter regime.
An
$A_{1g}+i B_{1g}$ state,
which breaks time-reversal symmetry, occurs at
low temperatures in part of the phase diagram.
Implications for the
experiments in the iron pnictides are discussed.}

\begin{document}

\maketitle

\section{Introduction}
The discovery of high temperature superconductivity
in the iron pnictides \cite{Kamihara_FeAs,Zhao_Sm1111_CPL08,Cheng:08}
has spurred tremendous experimental and theoretical interest
in these systems.
While they are not Mott insulators, the
undoped iron pnictides are nonetheless ``bad metals''.
This fact has motivated the placement of these systems
in an intermediate coupling regime close to the boundary between
Mott localization and itinerancy \cite{Haule,Si,Si_NJP},
where the Coulomb interactions bring about
non-perturbative effects in the form of incipient lower and upper
Hubbard bands. Accordingly, the low-energy Hamiltonian contains
quasi-localized moments with $J_1-J_2$ superexchange
interactions \cite{Si,Yildirim,Yin,Ma,Fang,Xu,Si_NJP},
and it corresponds to an effective multi-band $t-J_1-J_2$ model
for the carrier-doped systems.
Importantly, $J_1 \sim J_2 > J_1/2$ \cite{Si,Dai,Yildirim,Yin,Ma}
so that the system not only has a $(\pi,0)$
antiferromagnetic order,
as seen experimentally \cite{Cruz}, but also exhibits
strong magnetic frustration.
The incipient Mott picture is supported by the
observations
of the Drude-weight suppression \cite{Qazilbash,Hu,Si_natphys},
as well as the
temperature-induced spectral-weight transfer \cite{Hu,Yang,Boris}.
In addition, the quasi-localized moments are supported by the
inelastic neutron scattering experiments; Ref.~
\cite
{Zhao}, for instance,
observed zone-boundary spin waves and showed that $J_1$ is indeed
comparable to $J_2$.
(By contrast, Fermi surface in the magnetically ordered state
does not directly probe the strength of electron
correlations \cite{Si_NJP}.)
Alternatively, perturbative treatments \cite{Dong,Kuroki}
of the Coulomb interactions
have been used to study the magnetism of the iron pnictides.
The weak-coupling approaches, mostly based on spin fluctuations,
have been extensively used to address the
superconductivity \cite{Kuroki,Mazin,Tesanovic,Chubukov,Wang,Graser}.
By contrast, strong-coupling studies of superconductivity
have
been more limited \cite{Moreo,Seo,WChen,Berg}.

Experimentally, the pairing symmetry
in the iron pnictides has
remained inconclusive. The angle resolved photoemission\cite{Ding,Kondo}
(ARPES) and the Andreev spectroscopy \cite{TChen} results suggest a nodeless
gap. In contrast the nuclear magnetic resonance\cite{Nakai} (NMR) and some
penetration depth\cite{Gordon} measurements suggest a nodal
gap.
The experimental results seem to vary among pnictide
compounds. As an example, the P-doped BaFe$_2$As$_2$ appears to
have nodal gaps \cite{Matsuda,Stewart}
and this is in strong contrast with its K-doped counterpart.

In this Letter we address the pairing in the pnictides from
the incipient Mott approach, and treat the competing
pairing channels on an equal footing.
We are particularly motivated to consider the effect of the
$J_1 \sim J_2$ magnetic frustration,
and show that
it leads to a
quasi-degeneracy among several paring
states.

The Hamiltonian is given by
\begin{eqnarray}
H=H_t+H_{J1}+H_{J2} .
\end{eqnarray}
The kinetic part is
$H_t=-\sum_{i<j,\alpha,\beta,s} t_{ij}^{\alpha \beta}
c^{\dagger}_{i\alpha s}c_{j\beta s}+h.c.
-\mu \sum_{i,\alpha}n_{i\alpha}$,
where $c^{\dagger}_{i\alpha s}$
creates an electron at site $i$, with orbital $\alpha$ and spin
projection $s$; $\mu$ is the chemical potential and
$t_{ij}^{\alpha \beta}$ the hopping matrix.
The nearest-neighbor (n.n., $\langle ij\rangle$)
and next-nearest-neighbor (n.n.n., $\langle \langle ij \rangle \rangle$)
exchange interactions are,
respectively,
$H_{J1}= \sum_{\langle ij\rangle,\alpha,\beta} J_{1}^{\alpha \beta}
\left(\vec{S}_{i\alpha}\cdot \vec{S}_{j\beta}-\frac{1}{4}n_{i\alpha} n_{j\beta}\right)$,
and
$H_{J2}=\sum_{\langle \langle ij\rangle \rangle,\alpha,\beta} J_{2}^{\alpha \beta}
\left(\vec{S}_{i\alpha}\cdot \vec{S}_{j\beta}-\frac{1}{4}n_{i\alpha} n_{j\beta}\right)$.
Here,
$\vec{S}_{i\alpha}=\frac{1}{2}\sum_{s,s^{'}}c^{\dagger}_{i\alpha s}
\vec{\sigma}_{ss^{'}}c_{i\alpha s^{'}}$
and $n_{i \alpha}=\sum_{s}c^{\dagger}_{i\alpha s}c_{i\alpha s}$,
with $\vec{\sigma}$ representing the
Pauli matrices operating on the spin indices.
The above Hamiltonian is augmented by the appropriate
occupancy constraint for the fermions.
Note that, while the nearest-neighbor interaction has been evidenced to
be spatially anisotropic in the $(\pi,0)$ collinear antiferromagnetic
state \cite{Yin,Zhao},
it is expected to be isotropic in the tetragonal paramagnetic
phases and this is consistent with spin dynamical
measurements \cite{Diallo,Lester}.

\section{Two-orbital model}
We will first consider a two orbital model,
retaining only the $d_{xz} (\alpha=1)$ and $d_{yz} (\alpha=2)$
orbitals,
and later we consider a five orbital model to better address the fermiology \cite{Kuroki,PLee,Graser}.
We
further assume $J_{i}^{\alpha , \beta}
=J_i\delta_{\alpha,\beta}$, and $J_i>0$,
and
will subsequently
address the
role of interorbital exchange couplings.

Under the tetrahedral point group symmetry transformations ($D_{4h}$),
$d_{xz}$ and $d_{yz}$ orbitals transform respectively as $x$ and $y$
coordinates. The kinetic energy part of the Hamiltonian is invariant
under all point group symmetry operations ($A_{1g}$) and has the
following 
form in the extended Brillouin zone,
$H_{0}=\sum_{{\mathbf{k}},s}\psi^{\dagger}_{{\mathbf{k}}s}
\left[\xi_{{\mathbf{k}}+}\tau_0+\xi_{{\mathbf{k}}-}\tau_{z}
+\xi_{{\mathbf{k}}xy}\tau_{x}\right]\psi_{{\mathbf{k}}s}$,
where $\mathbf{k}=(k_x,k_y)$ and $\psi^{\dagger}_{{\mathbf{k}}s}
=(c^{\dagger}_{{\mathbf{k}}1s},c^{\dagger}_{{\mathbf{k}}2s})$.
The identity and Pauli matrices ($\tau_0$, $\tau_i$)
operate on the orbital indices, and $\xi_{{\mathbf{k}}+}
=-(t_1+t_2)(\cos k_x+\cos k_y)-4t_3\cos k_x \cos k_y-\mu$,
$\xi_{{\mathbf{k}}-}=-(t_1-t_2)(\cos k_x-\cos k_y)$,
$\xi_{{\mathbf{k}}xy}=-4t_4\sin k_x \sin k_y$ are respectively $A_{1g}$, $B_{1g}$, $B_{2g}$ functions.
The band dispersion relations $\mathcal{E}_{{\mathbf{k}}\pm}
=\xi_{{\mathbf{k}}+}
\pm \sqrt{\xi_{{\mathbf{k}}-}^2+\xi_{{\mathbf{k}}xy}^2}$, give rise to two electron pockets
at ${\mathbf{k}}=(\pi,0)$ and $(0,\pi)$, and two hole pockets
at ${\mathbf{k}}=(0,0)$ and
$(\pi,\pi)$.
The
carrier doping
$\delta=|\sum_{\alpha}n_{i\alpha}-2|$.
Mostly we use the minimal tight-binding model of Ref.~\cite{Raghu}, $t_1=-t$, $t_2=1.3t$, $t_3=t_4=-0.85t$
obtained from a fitting of the LDA bands.

\begin{figure}[t!]
\centering
\subfigure[]{
\includegraphics[scale=0.295]{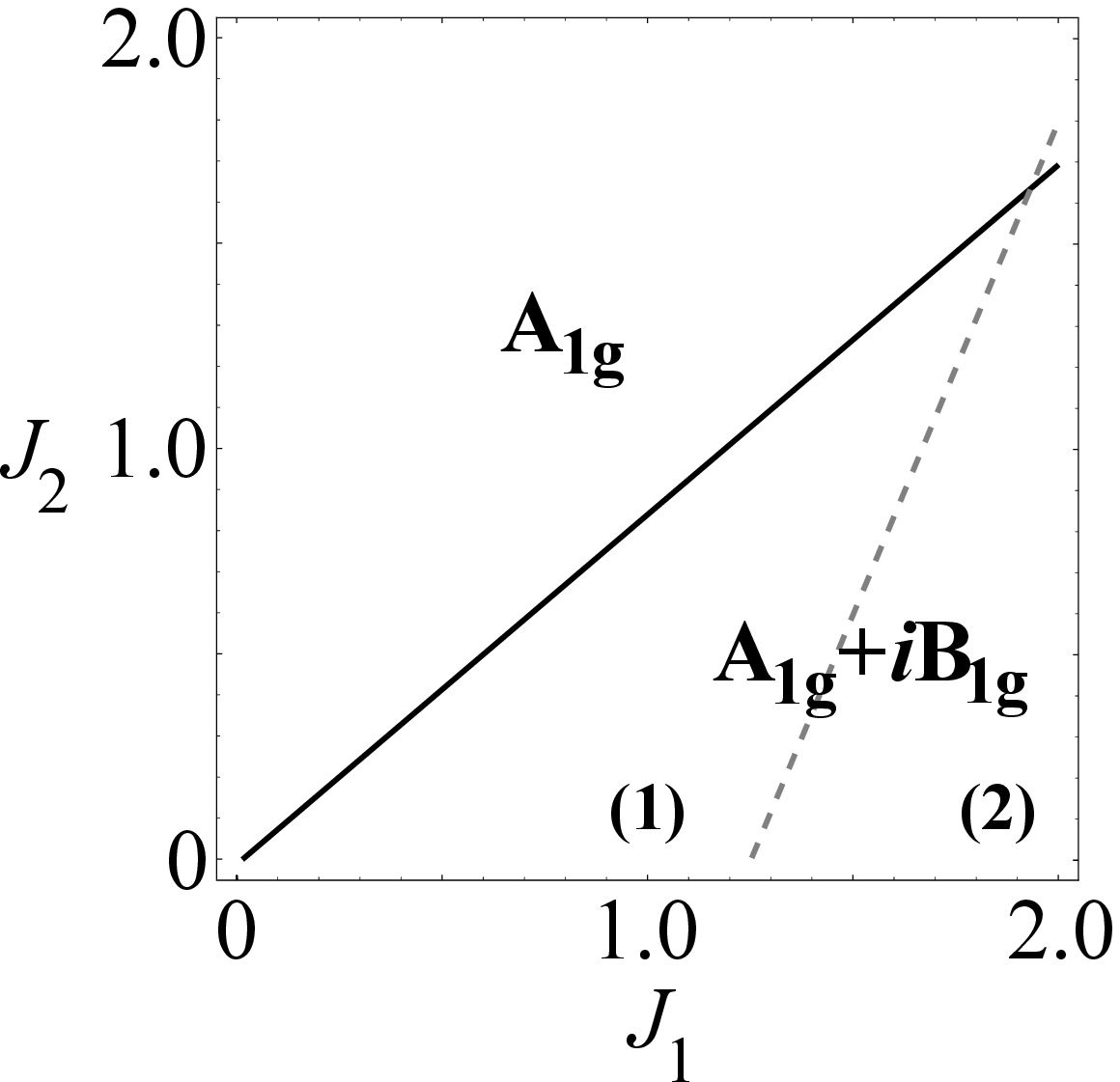}
\label{fig:subfig1}
}
\subfigure[]{
\includegraphics[scale=0.28]{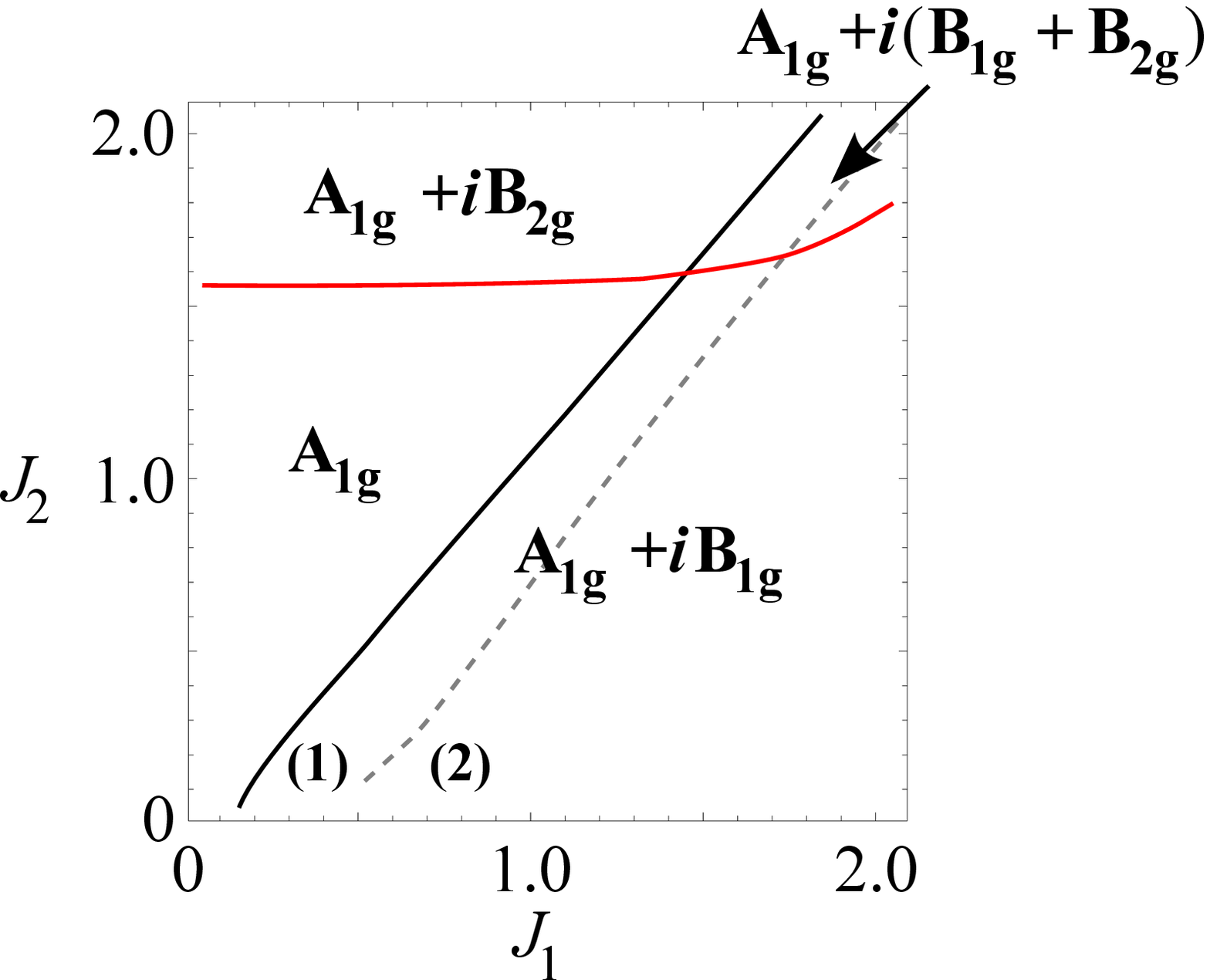}
\label{fig:subfig2}
}
\label{fig:subfigureExample}
\caption[]{Zero temperature phase diagrams of (a) a two orbital model
and (b) a five-orbital model, both for electron doping $\delta=0.14$. The onset of $B_{1g}$ and $B_{2g}$ phases are respectively marked by black and red solid lines. The dotted line characterizes a cross-over between $s_{x^2y^2}^{A_{1g}}$
[region (1)] and $s_{x^2+y^2}^{A_{1g}}$ [region (2)] as the dominant component
in the $A_{1g}$ pairing. 
}
\end{figure}

\subsection{Magnetic frustration and degeneracy of pairing states}
We start from the case with a vanishing kinetic energy,
in order to highlight the connection between magnetic frustration
and enhanced degeneracy of pairing states.
We define
the intra-orbital spin-singlet pairing
operators $\Delta_{{\mathbf{e}},\alpha \alpha}
=\langle c_{i\alpha\uparrow}c_{i+{\mathbf{e}}\alpha \downarrow}
-c_{i\alpha\downarrow}c_{i+{\mathbf{e}}\alpha \uparrow}\rangle/2$,
where ${\mathbf{e}}=\hat{x},\hat{y},\hat{x}\pm\hat{y}$.
Without the kinetic term, the problem decouples in the orbital
basis and we can drop the orbital indices. 
When $J_1$ dominates,
two degenerate pairing states $s_{x^2+y^2}$ and $d_{x^2-y^2}$,
respectively defined by the
pairing functions $g_{x^2\pm y^2,\mathbf{k}}=\cos k_x\pm\cos k_y$
are naturally favored. In real space, they respectively
correspond to $\Delta_x=\pm \Delta_y=\Delta_0$, with
$\Delta_{x+y}=\Delta_{x-y}=0$.
These $s_{x^2+y^2}$ and $d_{x^2-y^2}$ states are degenerate
because the symmetry operation $c_{m\hat{x}+n\hat{y}}\to
e^{i(2m+1)\pi/2}c_{m\hat{x}+n\hat{y}}$ transforms them
into each other.

When $J_2$ dominates, the $s_{x^2y^2}$ and
$d_{xy}$ states,
respectively defined by the
pairing functions
$g_{x^2y^2,\mathbf{k}}=\cos(k_x+k_y)+ \cos(k_x-k_y)$
and $g_{xy,\mathbf{k}}=\cos(k_x-k_y)-\cos(k_x+k_y)$,
are preferred, and they are degenerate. In real space, they correspond to
$\Delta_{x+y}=\pm \Delta_{x-y}=\Delta_0$,
with $\Delta_x=\Delta_y=0$. The $s_{x^2y^2}$ and $d_{xy}$ states transform into each other
by the following symmetry operation:
We break the square lattice into two interpenetrating
sublattices;on the even sublattice ($m+n=even$),
$c_{m\hat{x}+n\hat{y}}\to e^{i(m+n+1)\pi/2}c_{m\hat{x}+n\hat{y}}$,
and on the odd sublattice,
$c_{m\hat{x}+n\hat{y}}\to e^{i(m-n+1)\pi/2} c_{m\hat{x}+n\hat{y}}$.

As we tune the ratio $J_2/J_1$, we expect a level crossing in the
magnetically frustrated regime, $J_2 \sim J_1$. We can then anticipate that magnetic frustration
promotes an enlarged degeneracy among the
$s_{x^2y^2}$, $s_{x^2+y^2}$, $d_{xy}$, and $d_{x^2-y^2}$
pairing states.


\subsection{The effect of the kinetic energy}
When the kinetic term is incorporated, it lifts the exact degeneracies of the paired states discussed above.
We study the full problem using a mean-field decoupling \cite{Kotliar}
of the two-band $t-J_1-J_2$ model.
To set the stage,
we note that
the $D_{4h}$ point group symmetry operations allow
the following four classes of pairing states for an orbitally
diagonal $J_1-J_2$ model \cite{Zhou}:
(i) $\mathrm{A_{1g}}:[s_{x^2+y^2}^{A_{1g}}g_{x^2+y^2,\mathbf{k}}
+s_{x^2y^2}^{A_{1g}}g_{x^2y^2,\mathbf{k}}]\tau_0
+d_{x^2-y^2}^{A_{1g}}g_{x^2-y^2,\mathbf{k}}\tau_z$;
(ii) $\mathrm{B_{1g}}:d_{x^2-y^2}^{B_{1g}}
g_{x^2-y^2,\mathbf{k}}\tau_0
+[s_{x^2+y^2}^{B_{1g}}g_{x^2+y^2,\mathbf{k}}
+s_{x^2y^2}^{B_{1g}}g_{x^2y^2,\mathbf{k}}]\tau_z$;
(iii) $\mathrm{A_{2g}}: d_{xy}^{A_{2g}}g_{xy,\mathbf{k}}\tau_z$;
and (iv) $\mathrm{B_{2g}}: d_{xy}^{B_{2g}}g_{xy,\mathbf{k}}\tau_0$.
Each pairing channel will have different symmetry depending
on whether it is associated with $\tau_0$ or $\tau_z$ in the orbital
space; this distinction is denoted by the superscripts. The eight pairing amplitudes $s_{x^2+y^2}^{A_{1g}}$ etc. are linear combinations of eight intra-orbital pairing amplitudes $\Delta_{{\mathbf{e}},\alpha \alpha}$.

\subsection{Mean field theory}
We decouple
the exchange interactions
in the pairing channel.
In terms of
$\Psi^{\dagger}_{\mathbf{k}}
=(\psi^{\dagger}_{{\mathbf{k}}\uparrow}, \psi_{-{\mathbf{k}}\downarrow})$,
the Hamiltonian
becomes
\begin{eqnarray}
H_{mf}=\sum_{\mathbf{k}}\Psi_{\mathbf{k}}^{\dagger}
\left[\begin{array}{cc}\mathbf{h}_{\mathbf{k}}
& \mathbf{\Delta}_{\mathbf{k}}\\\mathbf{\Delta}^{\ast}_{\mathbf{k}}
& -\mathbf{h}_{\mathbf{k}}\end{array}\right]\Psi_{\mathbf{k}}
\label{H-mf}
\end{eqnarray}
where
$\mathbf{h}_{\mathbf{k}}=\xi_{{\mathbf{k}}+}\tau_0
+\xi_{{\mathbf{k}}-}\tau_{z}+\xi_{{\mathbf{k}}xy}\tau_{x}$, and $\mathbf{\Delta}_{\mathbf{k}}
=\mathrm{diag}[\Delta_{{\mathbf{k}},11},\Delta_{{\mathbf{k}},22}]$
is the orbitally diagonal gap matrix, and
$\Delta_{{\mathbf{k}},\alpha \alpha}
=\sum_{\mathbf{e}}J_{\mathbf{e}}\Delta_{{\mathbf{e}},\alpha \alpha}
\cos({\mathbf{k}}\cdot{\mathbf{e}}) $,
with
$J_{\mathbf{e}}=J_1$ for ${\mathbf{e}}=\hat{x},\hat{y}$,
and $J_{\mathbf{e}}=J_2$ for ${\mathbf{e}}=\hat{x}\pm \hat{y}$.
Diagonalizing
$H_{mf}$ we obtain the quasiparticle dispersion spectra $E_{{\mathbf{k}},\pm}$.
We determine
the pairing gap matrix by
minimizing the
ground state energy density
\begin{eqnarray}
f=\sum_{{\mathbf{e}},\alpha}\frac{J_{{\mathbf{e}}}}{2}|\Delta_{{\mathbf{e}},\alpha \alpha}|^2-\sum_{{\mathbf{k}},j=\pm}(E_{{\mathbf{k}},j}-\mathcal{E}_{{\mathbf{k}},j}),
\label{free_energy}
\end{eqnarray}
with respect to {\it all}
$\Delta_{{\mathbf{e}},\alpha \alpha}$.
The primary effect of the constraints
is to renormalize
the kinetic energy
via
(both
orbitally diagonal and off-diagonal)
$t \rightarrow t\delta/2$,
where $t$ is the kinetic energy scale.
Our results,
with an implicit treatment of the constraint through
a band renormalization,
remain largely unchanged when the
constraints are explicitly incorporated.

\begin{figure}[t!]
\centering
\includegraphics[scale=0.4]{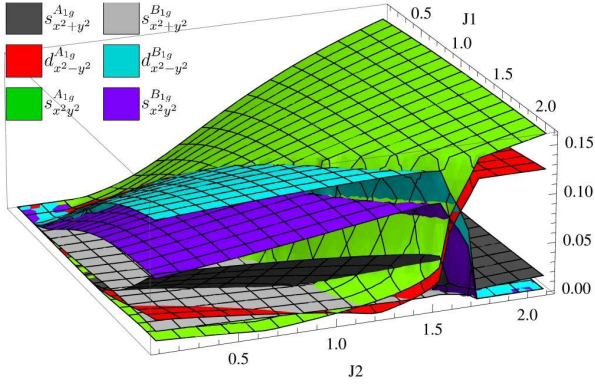}
\caption{(Color online) The amplitudes of different pairing
gap components of a two orbital model for electron doping $\delta=0.14$. For $J_2\gg J_1$ and $J_2\ll J_1$, $s_{x^2y^2}^{A_{1g}}$ ($s_{\pm}$), and $d_{x^2-y^2}^{B_{1g}}$ are respectively the dominant pairing channels.}
\label{fig:2}
\end{figure}

When
kinetic energy is absent,
the $s_{x^2+y^2}$ and $d_{x^2-y^2}$ states
are indeed degenerate,
and each has a ground state energy
$\approx -0.17 J_1$.
Likewise, the energy of either
$s_{x^2y^2}$ or $d_{xy}$ state is $\approx -0.17 J_2$, and all four
paired states become degenerate exactly at $J_1=J_2$,
as anticipated earlier.

We now turn to the results for the full problem
in the presence of the kinetic terms.
We show an illustrative zero-temperature
phase diagram for $0\leq J_1,J_2\leq 2t$,
in Fig.~\ref{fig:subfig1} corresponding to
an electron doping $\delta=0.14$.
$A_{1g}$ pairing exists in the entire $J_1-J_2$ plane.
In a sizable portion of the phase diagram,
the pairing is in the pure $A_{1g}$ class. In this
region $s_{x^2y^2}^{A_{1g}}$
is the dominant pairing channel and coexists  with
the subdominant $d_{x^2-y^2}^{A_{1g}}$ channel.
The onset of $B_{1g}$ pairing state is marked by
a solid line, which corresponds
to a second order phase transition.
$d_{x^2-y^2}^{B_{1g}}$ and $s_{x^2y^2}^{B_{1g}}$
are respectively the dominant and the subdominant
components of $B_{1g}$ phase.
In the large $J_1$ limit
(to the right of the dotted line),
$s_{x^2+y^2}^{A_{1g}}$ becomes the dominant
component of the $A_{1g}$ phase;
the dotted line represents a crossover. The phase diagram for hole doping is identical, but the onset of $B_{1g}$ pairing occurs for larger $J_1$.

\begin{figure}[t!]
\centering
\includegraphics[scale=0.16]{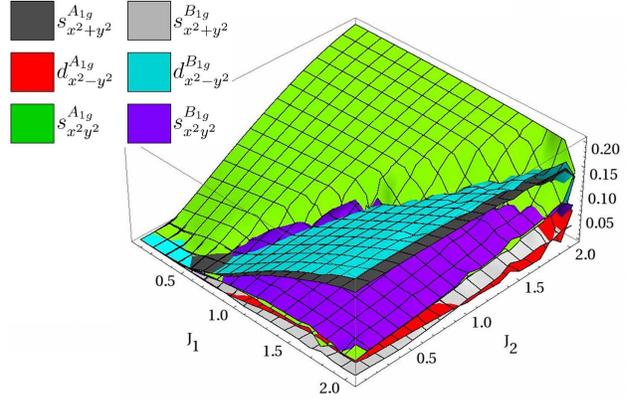}
\caption{(Color online) The pairing amplitudes for $d_{xz}$, $d_{yz}$ orbitals,
obtained from a five orbital model for $\delta=0.14$. For $J_2\gg J_1$, the dominant pairing channel is $s_{x^2y^2}^{A_{1g}}$ ($s_{\pm}$), and for $J_2\ll J_1$, $d_{x^2-y^2}^{B_{1g}}$ and $s_{x^2+y^2}^{A_{1g}}$ pairing channels are dominant and nearly degenerate pairing channels. Compared to the two-orbital model $s_{x^2+y^2}^{A_{1g}}$ channel is more competitive and becomes significant for relatively smaller value of $J_1$. 
}
\label{fig:3}
\end{figure}

The competition among different symmetry classes, and the nature
of the ground states are demonstrated in Fig.~\ref{fig:2},
which plot the pairing amplitudes as a function
of the coupling constants. The degeneracy between the $s_{x^2y^2}$
and $d_{xy}$ channels, occurs at very large $J_2$ limit and is not showed in Fig.~\ref{fig:2}.
For moderate values of $J_1, J_2$,
quasi-degeneracy among the pairing states
occurs in the magnetically-frustrated region corresponding to
$J_2 \sim J_1$.
In this region, Fig~\ref{fig:2} shows that the
weights of the $s_{x^2y^2}^{A_{1g}}$ and $d_{x^2-y^2}^{B_{1g}}$
components are comparable, and those of $d_{x^2-y^2}^{A_{1g}}$ and
$s_{x^2y^2}^{B_{1g}}$ are close-by.
This quasi-degeneracy also underlies
the phase diagram shown in
Fig.~\ref{fig:subfig1}.
By contrast, for moderate values of $J_2$ and with $J_2 \gg J_1$,
$s_{x^2y^2}^{A_{1g}}$
dominates over $s_{x^2y^2}^{B_{1g}}$, $d_{xy}^{A_{2g}}$, $d_{xy}^{B_{2g}}$
channels; likewise, for moderate values of $J_1$ and with $J_1 \gg J_2$,
$d_{x^2-y^2}^{B_{1g}}$ dominates
over $d_{x^2-y^2}^{A_{1g}}$, $s_{x^2+y^2}^{A_{1g}}$,
$s_{x^2+y^2}^{B_{1g}}$ channels.

At the $T=0$ limit we study, the co-existing pairing channels
will lock into a definite phase. Consider first the case
of coexisting $s_{x^2y^2}^{A_{1g}}$
and $d_{x^2-y^2}^{A_{1g}}$. We define
$\Delta_{\mathbf{k}}
= s_{x^2y^2}^{A_{1g}}g_{x^2y^2,\mathbf{k}}\tau_0+d_{x^2-y^2}^{A_{1g}}
g_{x^2-y^2,\mathbf{k}}\tau_z$, and use $\phi$ to denote the relative phase.
Inserting this into Eq.~(\ref{free_energy}), we find that
$\phi=\pi$ corresponds to the ground-state energy minimum, with
$\partial f/\partial \phi =0$ and $\partial f/\partial \phi > 0$
($\phi=0$ is a ground-state energy maximum).
The $A_{1g}$ gap is therefore real. Moreover, the relative
minus sign associated with $\phi=\pi$ implies that
the pairing
function, in the band (as opposed to the orbital) basis,
is $s_{x^2y^2}^{A_{1g}}g_{x^2y^2,\mathbf{k}}\tau_0
-d_{x^2-y^2}^{A_{1g}}g_{x^2-y^2,\mathbf{k}}(\xi_{\mathbf{k}-}\tau_z-\xi_{{\mathbf{k}}xy}\tau_x)
(\xi_{\mathbf{k}-}^2+\xi_{\mathbf{k}xy}^2)^{-1/2}$. Hence intraband pairing function
changes sign between the hole and electron pockets near
$\mathbf{k}=(0,0)$ and
$\mathbf{k}=(\pi,0), (0,\pi)$ respectively. A similar argument shows
that the relative phase between $d_{x^2-y^2}^{B_{1g}}$ and
$s_{x^2y^2}^{B_{1g}}$ is $\pi$, and the gap changes sign between
the two electron pockets near $\mathbf{k}=(\pi,0), (0,\pi)$ respectively.

Consider next the case of coexisting $s_{x^2y^2}^{A_{1g}}$ and
$d_{x^2-y^2}^{B_{1g}}$. A similar analysis shows that, this time,
$\phi=\pi/2$ corresponds to the ground-state energy minimum
(while $\phi=0,\pi$ represent
energy maxima).
The resulting $A_{1g}+iB_{1g}$ phase describes
the state to the right of the solid line in
Fig.~\ref{fig:subfig1}.
This phase simultaneously breaks time reversal and four-fold rotational
symmetries, but preserves the combination of the two symmetries.
Such a state also occurs in a phenomenological Landau-Ginzburg
theory\cite{Lee}.

The role of fermiology can be clearly illustrated in the linearized
gap approximation. For \emph{a set of small Fermi pockets} at $(0,0)$,
$(\pi,0)$ and $(0,\pi)$, compared to $g_{xy,{\mathbf{k}}}$, $g_{x^2y^2,{\mathbf{k}}}$ has larger overlap with the pairing kernel.
Thus $s_{x^2y^2}^{A{1g}}$ and $s_{x^2y^2}^{B_{1g}}$ gaps have higher $T_c$'s compared to $d_{xy}^{B_{2g}}$ and $d_{xy}^{A_{2g}}$ gaps;
they become degenerate only in the large $J_2$ limit.
Similar reasoning shows that, unless a threshold value for $J_1$ is
exceeded, $d_{x^2-y^2}^{A_{1g}}$ and $d_{x^2-y^2}^{B_{1g}}$ gaps have
higher $T_c$'s compared to the $s_{x^2+y^2}^{A_{1g}}$ and
$s_{x^2+y^2}^{B_{1g}}$ gaps.
Related observations were made by
Seo {\it et al.} \cite{Seo}. If we consider a band structure that produces large pockets\cite{Moreo},
$d_{xy}^{B_{2g}}$ replaces $s_{x^2y^2}^{A_{1g}}$ as the
dominant pairing state for $J_2>J_1$, and we observe a competition between $d_{xy}^{B_{2g}}$ and $d_{x^2-y^2}^{B_{1g}}$ pairing states.
Note that the magnetic oscillation and ARPES measurements
suggest small sizes of the Fermi pockets for the iron pnictides,
and this should make our $A_{1g}$ and $A_{1g}+iB_{1g}$ phases more
feasible.

\subsection{Effects of inter-orbital exchange couplings}
We stress that, while the detailed nature of the lattice symmetry and
orbitals for the single-particle energy dispersion is important for
a proper description of the Fermi surface, its counterpart for the
exchange interactions is not obviously so. In the absence of a detailed
knowledge about such structure in the exchange interactions, we have
considered the simplest description and focused on the associated
properties that are qualitative and robust. For instance,
we have so far considered intra-orbital exchange interactions.
Inclusion of inter-orbital super-exchange interactions does not
change the phase diagram. Now the $A_{1g}$ phase will have a small
inter-orbital $d_{xy}^{A_{1g}}$ component and the $B_{1g}$ phase
will remain unchanged.
Hund's coupling may lead to on-site triplet pairing, but this
is prevented by the onsite Coulomb repulsion that is built in
our model. Inter-site triplet pairing may arise if the Hund's coupling
is comparable to Coulomb repulsion \cite{WChen},
but this is unlikely on
general or {\it ab initio} grounds; it is also unlikely on empirical grounds
since experimental evidence has so far been overwhelming for
singlet pairing.
Using a multi-orbital Hubbard model
as their starting points, both the strong coupling
calculations of Ref.~\cite{WChen} and the
RPA calculations of Ref.~\cite{Kuroki}
find that a moderate $J_H$ enhances repulsive inter-electron-pocket
pair scattering and leads to stronger $B_{1g}$ pairing.
Thus $J_H$ can only enhance the $A_{1g}+iB_{1g}$ part of the
phase diagram.
We have preferred to consider the multiband
$t-J_1-J_2$ model, since $J_1$ and $J_2$ can be more readily connected with
the magnetic frustration physics.

\section{Five-orbital model}
To understand the robustness of our two-band
results, against the inclusion of additional bands,
and better address the fermiology
we have considered a five-band model with the kinetic terms
according to Ref.~\cite{Graser}. To capture the
important results within a simple model of interaction, we again choose a $J_1-J_2$ model with
$J_{i}^{\alpha \beta}=J_i\delta_{\alpha \beta}$. Now the general intra-orbital pairing matrix has
the form $\mathbf{\Delta}_{\mathbf{k}}=\sum_{a}\mathrm{diag}[\Delta^{a}_{{\mathbf{k}},11},
\Delta^{a}_{{\mathbf{k}},22},\Delta^{a}_{{\mathbf{k}},33},\Delta^{a}_{{\mathbf{k}},44},
\Delta^{a}_{{\mathbf{k}},55}]$, where the index $a$ corresponds to $s_{x^2+y^2}$, $d_{x^2-y^2}$, $s_{x^2y^2}$
and $d_{xy}$ symmetries.
The results of minimizing the free energy with respect to
all the twenty
complex pairing amplitudes
are given in an illustrative phase diagram Fig.~\ref{fig:subfig2},
and in Fig.~\ref{fig:3},
which shows the competition among the pairing amplitudes
for $xz$ and $yz$ orbitals.
We find that the competition between $A_{1g}$ and $B_{1g}$ pairings is a robust effect. In contrast to the two band case $d_{x^2-y^2}^{A_{1g}}$ amplitude is reduced and $s_{x^2+y^2}^{A_{1g}}$ amplitude is enhanced in the entire phase diagram.
Indeed, for the magnetically-frustrated $J_1\sim J_2$ region,
the nodal $A_{1g}$ $s_{x^2+y^2}$ and
$B_{1g}$ $d_{x^2-y^2}$ states are competitive against
the nodeless $A_{1g}$ $s_{x^2y^2}$ state.
Compared to the two band case the $B_{2g}$ pairing occurs for smaller $J_2$, but it still occurs in the limit when $J_2$ is bigger than the kinetic energy scale. The other three orbitals also
demonstrate similar competition among $s_{x^2y^2}$, $s_{x^2+y^2}$, $d_{x^2-y^2}$, and $d_{xy}$ pairings.

\section{Experimental implications of quasi-degenerate pairing channels}
Some of the weak-coupling studies \cite{Kuroki, Graser} have also
indicated a competition between
various pairing
states.
However, the weak-coupling approaches are typically restricted
to an instability analysis of the linearized gap equations.
By contrast, the strong-coupling
approach used here has the advantage of readily considering
the non-linear gap equations.
Our non-linear analysis is important in bringing out
the connection between the quasi-degeneracy in the
pairing channel and the $J_1 \sim J_2$ magnetic frustration.
 With only one
requirement ($J_1 \sim J_2$), which is linked to magnetic frustration
and has been supported by both theoretical
considerations \cite{Si,Dai,Yildirim,Yin,Ma}
and inelastic magnetic experiments \cite{Zhao,Diallo},
our result provides a parameter-insensitive mechanism
for the near degeneracies of various paring states.
Moreover, our non-linear analysis
is also essential in reaching the conclusion
({\it cf.}
Figs.~\ref{fig:2} and \ref{fig:3})
that, for $J_1 \sim J_2$,
the pairing amplitudes for
several competing nodeless and nodal pairing
channels are comparable.
This last result
is
particularly
important for experiments in the iron pnictides.
Indeed,
our result
has anticipated
the recent
experimental findings \cite{Matsuda,Stewart} that
pnictide superconductors with nodeless or nodal gaps
have a comparable maximum $T_c$. Our $s_{x^2y^2}^{A_{1g}}$ ($s_{\pm}$), $s_{x^2+y^2}^{A_{1g}}$  states appear to be consistent with ARPES measurements that find full gap at hole pockets. However in contrast to $s_{x^2y^2}^{A_{1g}}$, which is fully gapped on all the Fermi pockets as long as the pockets are not too large, $s_{x^2+y^2}^{A_{1g}}$ state has nodes on the electron pockets. This may be the reason for seeing nodal behavior in P-doped 122 compounds in contrast to the fully gapped behavior in K-doped 122 compounds.

To summarize, we have shown that magnetic frustration effects lead to
quasi-degeneracy among different pairing states.
With the bandstructures appropriate for the iron pnictides,
an extended $A_{1g}$ state is the dominant pairing state,
but this state contains both $s_{x^2y^2}$ and
$s_{x^2+y^2}$
components. Moreover, a $B_{1g}$ state has a close-by
ground state energy.
The quasi-degeneracy makes it likely that
the iron pnictides of different material families,
or
different
dopings, have different superconducting states.
Our detailed phase diagram contains a low-temperature phase
with time-reversal-symmetry breaking, which
also can be tested
by future experiments.

\acknowledgments
We thank E. Abrahams, B. A. Bernevig, and A. Nevidomskyy
for useful discussions, and acknowledge the support of the NSF
Grant No. DMR-0706625,
the Robert A. Welch Foundation
Grant No. C-1411,
and the W.\ M.\ Keck Foundation.

\end{document}